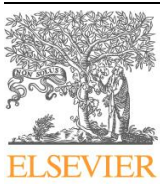

# Image Super-Resolution Using TV Priori Guided Convolutional Network

Bo Fu[a]∗, Yi Li[a], Xiang-hai Wang[a]

[a] *School of Computer and Information Technology, Liaoning Normal University, No 850, Huanghe Road, Dalian, 116029,China.*

## ABSTRACT

We propose a TV priori guided deep learning method for single image super-resolution (SR). The new up-sample method based on TV priori information, new learning algorithm and neural networks architecture are embraced in our TV Priori Guided Convolutional Neural Network which directly learns an end-to-end mapping between the low/high-resolution images. There are three aspects of the innovation of our algorithm. First, in many deep learning based super-resolution algorithms, an input image is up-sampled via bicubic interpolation before they fed into the network. So we define a set of discrete TV council's templates to extract TV priori information. Second, we put TV priori information into non-local regression frame, and propose a non-local TV interpolation; Then, non-local TV interpolated images are fed into the network. We conduct a number of experiments to evaluate the effectiveness TV priori guided CNN.

---

∗ Corresponding author. Tel.: +86-411-82155588; fax: +0-000-000-0000; e-mail: fubo@lnnu.edu.cn

## 1. Introduction

Currently, image super-resolution reconstruction(SR) is a fairly active research field , as it provides solutions overcoming low-resolution limitations from cell phone imaging to remote sensing imaging to medical imaging. Low-quality images are usually caused by low-cost sensor capture, narrow bandwidth transfer and bad light interference. However, when low-quality images are widely used for high-definition display, visual analysis and recognition tasks[1,2,3,37] they tend to be displayed in higher resolution versions. Although this problem has been studied, image super-resolution is still a highly challenging task that estimates a high-resolution(HR) version from the content of low-resolution images(LR). From Figure 1, it can be seen that a small image is firstly enlarged into a big scale image. For one point in small image, three points need to be estimated in reconstructed image. This problem is inherently ill-posed since it is difficult to obtain abundant information including content, light condition and regression(e.g. blur, noise, etc.) from images. This under-determined inverse problem will lead to a unique solution, though people have tried to design optimization methods and expression models. By now, people only can do everything possible to dig prior knowledge to build model which observes mapping relationship from low-resolution images to high-resolution images. Then, super-resolution reconstruction algorithm can be designed based on this generation model to reproduce high-resolution images.

A class of SR approaches use simple linear function to up-sample, such as Bilinear or Bicubic and S-Spline[4-5]. However, this kind of reconstruction-based SR algorithms always work badly when a larger magnification factor is desired. At the same time, the quality of small images is also closely related to the quality of algorithms[6]. Some researches tended to find more powerful local descriptive model, Dai et al. used image patches as local represent model, and reconstruct descriptors between background/foreground [7]. Sun et al. added the gradient prior information into local image structure [8]. Such class of SR approaches usually tended to introduce more complex structures (patches, etc.) and auxiliary judgment information(gradient, edge detection information, etc.). But most of them are not learning-based, i.e. these approaches fix an invariant set of parameters of models for different content of image data. So over-smoothing or burr phenomena often happen when these approaches process natural images with different texture contents.

The second class of SR approaches is learning-based, which tries to use machine techniques to learning parameters of representative models. Among them, some approaches dig internal similarities from the same LR image[9-11], then use regression methods to restore those blank points. These restoration techniques have been discussed, improved and applied widely in image denoising field [12-15]. Some other approaches are example-based methods. They dig similarities from external image pairs or image set, for example they trained mapping functions from low/high resolution image pairs [16-18]. Since the above methods can learn the parameters of the model through a large number of learning samples during the learning process, they generally obtain more excellent results. At the same time, due to too many data acceleration strategy or pre-treatment (such as sparse-coding, Low-Rank Kernel, clustering and PCA) often needs to be considered[19-23,37,38].

Recently, deep learning methods provide a new solution idea for SR methods which learn a mapping between the LR and HR images. Dong et al. firstly addressed a CNN to learn a mapping from LR to HR and named Super-Resolution Convolutional Neural Network(SRCNN)[24]. SRCNN owns simple networks architecture but excellent results. Kim et al. first introduced the residual network for training very deeper super-resolution networks named (VDSR), which can handle multi-scales resolutions jointly in same network[25]. However, we find these networks still exists some limitations in terms of network architectures.

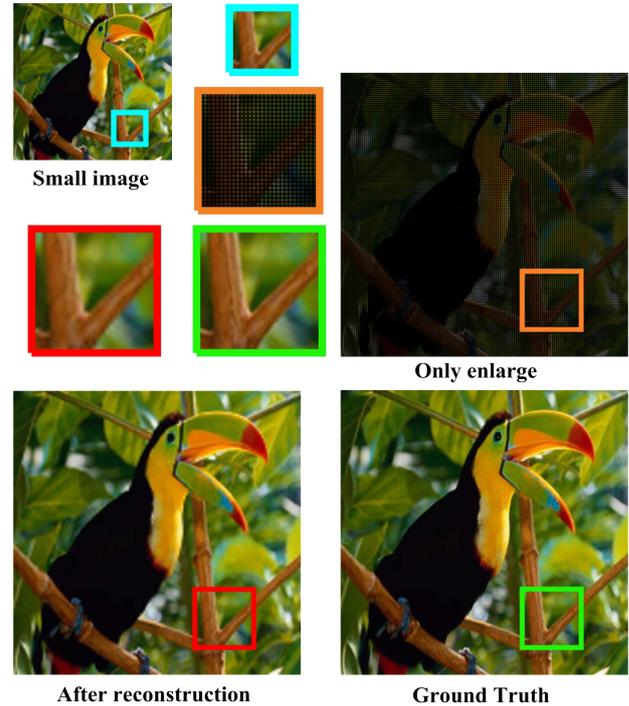

Fig 1. Small image, up-sampled image, SR image (bicubic) and Ground Truth image and their enlarged patch

First, most SR methods adopt Bicubic as pre-processing step, but traditional Bicubic interpolation is local linear function, so its reconstruction performance is not sensitive to minor architectural changes. Second, pre-processing layer of the neural network is not made the best of self-similarity.

In this work, we propose a TV priori information guided super-resolution reconstruction using. Aiming at two points of weakness mentioned above, we firstly adopt Maximal-Order B-splines as pre-processing step, and then defined a set of discrete TV templates to dig potential texture content priors; second, we use TV priori information into a non local self-learning framework. In addition, this TV priori information is embed to a integrated deep convolutional neural networks.

Overall, the contributions of this study are mainly inhere aspects: (1) we present a fully TV deep convolutional neural network for image super-resolution. The network directly learns an end-to-end mapping between   (2) non-local self-similarity learning will provide guidance for the design of the network structure.

## 2. Related work

### 2.1. Image Super-Resolution method

Early super-resolution approaches usually use interpolation techniques, and most of these methods are local. However, those methods exhibit limitations in predicting. Because linear kernel functions only can partial approximation due to limited amount of data. So example-based methods usually are adopt to improve the compatibility of the model with the data. Internal example-based methods exploit the self-similarity property from single image; it is widely used in image denoising field [15]. Glasner used priori information into a regression [10]. Some example-based methods are proposed, these studies vary on how to learn a compact dictionary or manifold space between low resolution and high solution[19,23,26,28]. Most of them use Bicubic as a pre-processing step.

After up-sampling step, amplified images by linear methods (like Bicubic) often lose some detail due to smoothing. Therefore, how to excavate prior information to guide image repair is a key issue. According to the image prior's information, image super resolution algorithms usually use prediction models, edge based methods, image statistical models as an auxiliary information for algorithm analysis. And then, example-based learning methods usually are adopted to estimate detailed parameters[10,18,26,27].

*2.2. Convolutional Neural Networks*

Recently, Convolutional neural networks (CNN) has led to dramatic improvements in SR. Dong et al. [24] first proposed a deep learning-based SR method, and they speeded up it[36]. Kim et al. [18] first introduced the residual network for training much deeper network architectures.

It can be seen, many deep learning based super-resolution algorithms, an input image is upsampled via bicubic interpolation before they fed into the network [24,36]. However, when image magnification factor is high, the effect of bicubic is often poor.

In this work, we change bicubic as a B-spline interpolation algorithm. And added a non-local TV priori step to enhance detail information.

## 3. TV Priori Information Guided Convolutional Neural Networks for Super-resolution.

*3.1. TV priori information extract*

TV model is proposed by Rudin firstly [29], a significant advantage of TV model is that it allows discontinuity points exist in variation function space $BV(\Omega)$ (it leads to an underlying sparse solution). Set $\Omega \subset \mathbf{R}^2$, $U \in L^1(\Omega)$ is a bounded open collection, it is usually assumed to be a Lipschitz domain. Assume U can be expressed as function u(i,j), and u is smooth. The TV express of u is:

$$\min_{u \in BV(\Omega)} TV[u] = \int_\Omega |\nabla u| dxdy, \quad \nabla u = (u_x, u_y) \quad (1)$$

Here, u satisfies the following constraint conditions:

$$\int_\Omega u(x,y)dxdy = \int_\Omega u_0(x,y)dxdy, \quad \frac{1}{|\Omega|}\int_\Omega (u(x,y) - u_0(x,y))^2 dxdy = \sigma^2 \quad (2)$$

Further, the document [30] defines 8 kinds of contour templates to discretize the contour stencils(CS) information of the integral region, CS is defined as:

$$(S^*[u])(k) := \sum_{m,n \in \mathbf{R}^2} S(m,n) |u_{k+m} - u_{k+n}| \quad (3)$$

For each pixel, they find aimed at the contour stencils which value of $(S^*[u])(k)$ is minimum. But, these templates only cover 8 directions and they are under the assumption that the function is smooth. And then we extend 8 directions to 3 types of directional templates including horizontal direction, vertical direction, and diagonal direction. Each direction including 8 contour stencils and corresponding discretization templates. We define discrete templates for the purpose of extract directional priori information. The specific definitions are given in Table 1.

We denotes $S_{k_d}^{(d)}$ as our extended contour templates, and upper corner mark $d$ represents 3 kinds of directional classes(i=1,

2, 3), under corner mark $k$ represents 8 directional contour templates(k=1,2,…8). Where $(S_{k_d}^{(d)} . *[u_{(i,j)}])$ is seen a discrete estimate value i.e. $(S_{k_i}^{(d)} . *[u_{(i,j)}]) \approx \|u_{k_i}^d\|_{TV(C^{(d)})}$. We can generate the corresponding contour stencil by minimum value of $(S_{k_d}^{(d)} . *[u_{(i,j)}])$. For a patch p, we denote corresponding contour stencils as **cs**.

We can get directional information using these TV contour stencils. Figure 2 shows directional priori information from a part of tested image, each block own two main directions.

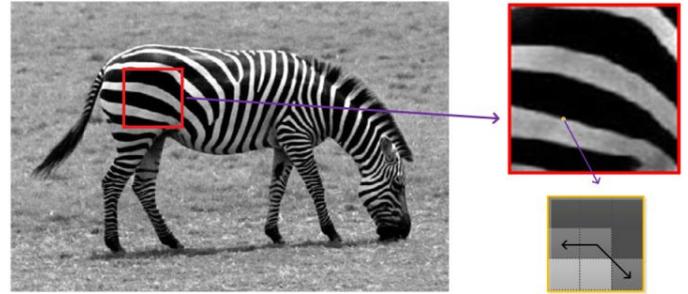

Fig 2. directional priori information from a block of tested image

*3.2. TV priori in non-local regression frame*

For a enhancing pixel (i,j) in patch P, we can get a set of most similar patches Q. We denote the contour stencil of P as $CS_p$, the contour stencil of Q as $CS_q$. We adopt contour stencils to build filter because they contains more priori information. In the regression framework, we can use these contour stencils to construct non local regression.

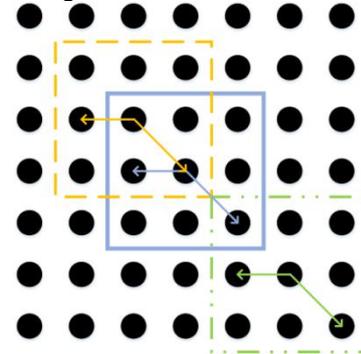

Fig 3. Patch searching similar patches in non-local scope

Let the restored gray value of the corrupted pixel be $\tilde{f}_{(i,j)}$, which is obtained by weighted averaging as follows:

$$\tilde{f}_{(kk)} = \sum_{\forall (q) \in R} w_{(CSp,CSq)} \cdot f_{(kk)} \quad (4)$$

Here, any $CSq$ belong to the similar patches **R**, $w_{(CSp,CSq)}$ is the weight according to the contour stencil $CS_p$ and $CS_q$. Weight $w_{(CSp,CSq)}$ is calculated using the similarity $s_{(CSp,CSq)}$ between each reference contour stencil and the target contour stencil, the corresponding formula is as follows:

$$w_{(CSp,CSq)} = \frac{s_{(CSp,CSq)}}{\sum_{1}^{mm} s_{(CSp,CSq)}} \quad (5)$$

Here, mm is the number of similar contour stencils. The specific definition of $s_{(CSp,CSq)}$ is as follow:

$$s_{(CSp,CSq)} = \frac{1}{e^{\frac{\|CSp-CSq\|^2}{\sigma}} \ln \sigma} \qquad (6)$$

Here, $\sigma$ is used to control strength of the filtering corrosion.

### 3.3 TV prior CNN

So far, all patches in set are fed in TV priori non-local to enhance. We input these patches into our networks. Our networks own three layers, in each layers, we define a set of filters and operators to generate mappings.

In the first layer, we represent these patches by a set of bases. The operation can by describe as follows:

$$F_{i(i\in[1,2,3])}(P) = \max(0, W_i * P + B_i) \qquad (7)$$

where W and B represent the bases filter set and biases respectively, size of filter is f1.

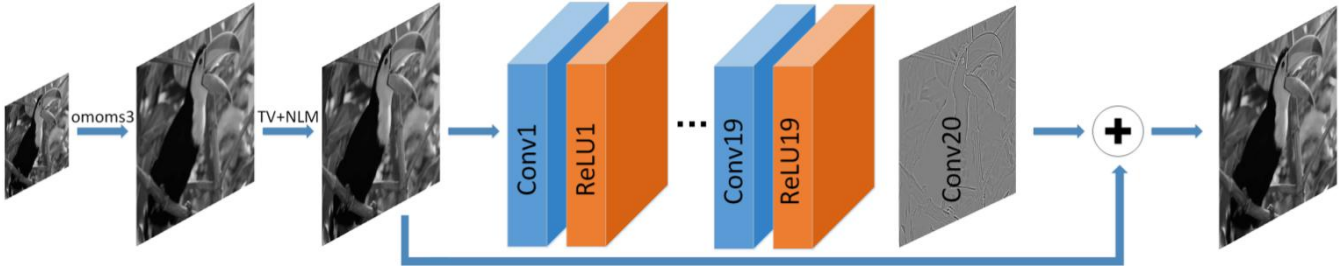

Fig 4. Structure of TV priori information guided CNN

We denote the number of bases filters is n, and filter's size is same to size of patch. So, each patch is applied n convolutions. The convolutional result of each patch is corresponding an n-dimensional feature map. Then we apply the Rectified Linear Unit (ReLU, max(0,x)) [24] on the filter responses to obtain more non-linearity.

## 4. Experiments

In this section, empirical results are shown. We compare our TV priori information network against the sate-of-the-art baseline algorithms.

### 4.1. Experiment setting

**Dateset** We employ a public training set from [20,31], which contains 91and 200 training images sets respectively. During testing, we use three different sets to evaluate our algorithm, i.e. set 5[32], set14[33] and BSD100[22]. Each image set also own corresponding Ground Truth sets.

**Baseline algorithm** we use seven existing algorithms as baselines, including 5 traditional algorithms and two neural network based one.

The traditional algorithms A+[22], SelfExSR[34], RFL[35], KK[18] and ANR[26]. The neural network based algorithms are SRCNN-s proposed in [24] and FSRCNN-s proposed in [36].

**Evaluation Metric** The peak signal-to-noise ratio (PSNR) is adopted to measure the objective performance of our algorithm. PSNR is defined as follows.

$$PSNR = 10\log_{10}(\frac{255^2}{MSE}) \qquad (8)$$

Here, MSE is Mean Square Error, it is defined in Formula 8.

### 4.2. Performance Comparison

Table2 shows comparisons results with several classical interpolations. It can be seen that our TV priori information guided CNN can obtains higher PSNR scores. The visual results are also compared in Figure 5. We compared image flowers, image foreman and image head, respectively.

**Table 2. Results of basic data comparison**

| Datasets | Evaluation Indexes | Bicubic | A+ | SelfExSR | RFL | KK | SRCNN-s | FSRCNN-s | Ours |
|---|---|---|---|---|---|---|---|---|---|
| Set5 | PSNR | 33.66 | 36.55 | 36.49 | 36.54 | 36.20 | 36.34 | 36.58 | **36.60** |
| Set5 | SSIM | 0.930 | 0.954 | 0.954 | 0.954 | 0.951 | 0.952 | 0.953 | **0.963** |
| Set14 | PSNR | 30.23 | 32.28 | **32.32** | 32.26 | 32.11 | 32.18 | 32.28 | 31.45 |
| Set14 | SSIM | 0.869 | 0.906 | 0.904 | 0.904 | 0.903 | 0.903 | 0.905 | **0.924** |

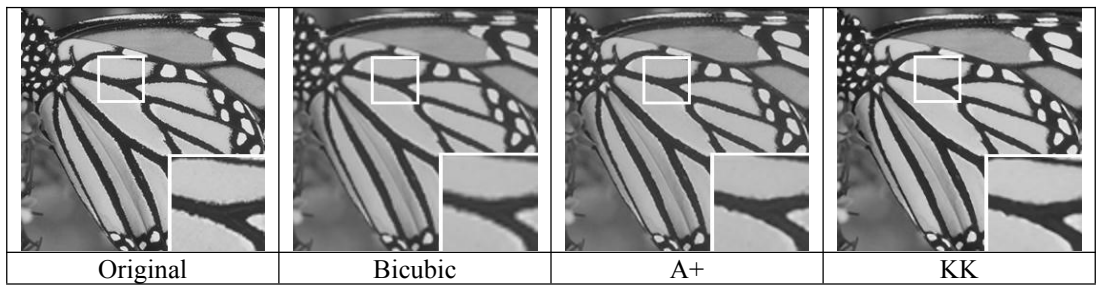

| Original | Bicubic | A+ | KK |

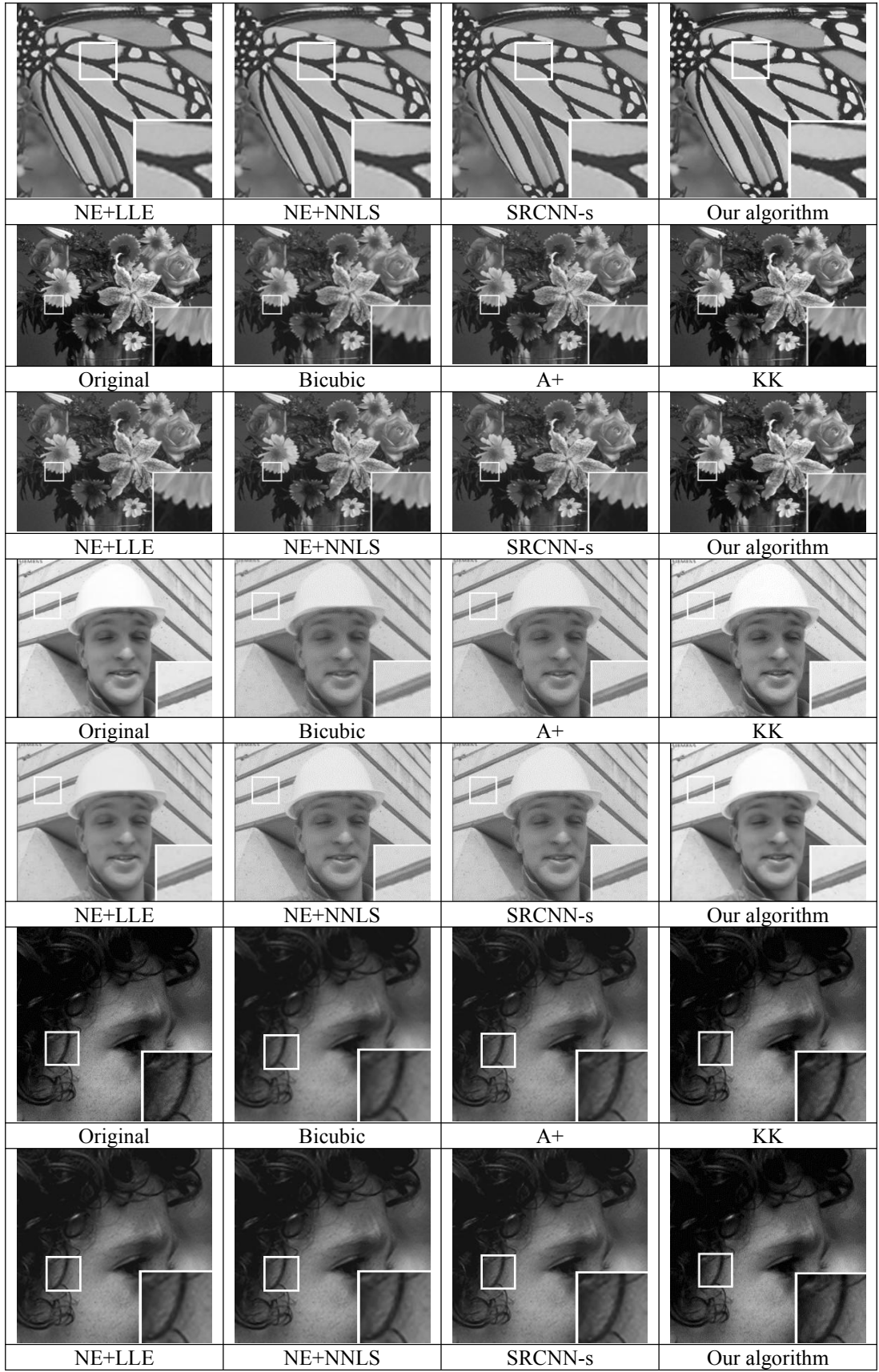

## 5. Conclusion

We propose a TV priori guided deep learning method for single image super-resolution (SR). The new up-sample method based on TV priori information, new learning algorithm and neural networks architecture are embraced in our TV Priori

Guided Convolutional Neural Network which directly learns an end-to-end mapping between the low/high-resolution images. There are three aspects of the innovation of our algorithm. First, in many deep learning based super-resolution algorithms, an input image is up-sampled via bicubic interpolation before they fed into the network.

**Acknowledgments**

This work is supported by the National Natural Science Foundation of China (NSFC) Grant Nos. 61702246, Liaoning Province of China General Project of Scientific Re-search No. L2015285, Liaoning Province of China Doctoral Research Start-Up Fund No. 201601243.

**References**

1. Y. Wang, X. Lin, L. Wu, W. Zhang, Q. Zhang, X. Huang. Robust subspace clustering for multi-view data by exploiting correlation consensus. IEEE Transactions on Image Processing, 24(11):3939-3949,2015.
2. L. Wu, Y. Wang, J. Gao, X. Li. Where-and-When to Look: Deep Siamese Attention Networks for Video-based Person Re-identification. IEEE Trans. Multimedia, 2018.
3. Y. Wang, L. Wu. Beyond low-rank representations: Orthogonal clustering basis reconstruction with optimized graph structure for multi-view spectral clustering. Neural Networks, 103:1-8, 2018.
4. Pascal Getreu. Linear Methods for Image Interpolation. Image Processing on Line , 2011 , 1:239-259.
5. R. Keys, Cubic Convolution Interpolation for Digital Image Processing, IEEE Transactions on Acoustics, Speech, and Signal Processing, 2003 , 29 (6) :1153-1160 .
6. S. Baker and T. Kanade, Limits on super-resolution and how to break them, IEEE Trans. Pattern Anal. Mach. Intell., vol. 24, no. 9, pp.1167–1183, Sep. 2002.
7. S. Dai, M. Han, W. Xu, Y. Wu, and Y. Gong, Soft edge smoothness prior for alpha channel super resolution, IEEE Conf. Comput. Vis. Pattern Class., 2007, pp. 1–8.
8. J. Sun, Z. Xu, and H. Shum, Image super-resolution using gradient profile prior, IEEE Conf. Comput. Vis. Pattern Recognit., 2008, pp. 1–8.
9. Freedman, G., Fattal, R.: Image and video upscaling from local elf-examples. ACM Transactions on Graphics 30(2), 12 (2011).
10. Glasner, D., Bagon, S., Irani, M.: Super-resolution from a single image. In: IEEE International Conference on Computer Vision. pp.349–356 (2009).
11 Huang, J.B., Singh, A., Ahuja, N.: Single image super-resolution from transformed self-exemplars. In: IEEE Conference on Computer Vision and Pattern Recognition. pp. 5197–5206 (2015)
12 Yang, J., Lin, Z., Cohen, S.: Fast image super-resolution based on in-place example regression. In: IEEE Conference on Computer Vision and Pattern Recognition. pp. 1059–1066 (2013).
13 L. Wu, Y. Wang, L. Shao. Cycle-Consistent Deep Generative Hashing for Cross-Modal Retrieval.arXiv preprint arXiv:1804.11013, 2018.
14 Y. Wang, L. Wu, X. Lin, J. Gao. Multiview spectral clustering via structured low-rank matrix factorization. IEEE Transactions on Neural Networks and Learning Systems, 2018.
15. Buades, A., Coll, B., Morel, J. M.: A non-local algorithm for image denoising. In: IEEE Computer Society Conference on Computer Vision and Pattern Recognition 2005, vol. 2, pp.60-65. IEEE (2005).
16. Freeman, W.T., Pasztor, E.C., Carmichael, O.T.: Learning low level vision. International Journal of Computer Vision 40(1), 25–47 (2000).
17. H. Chang, D.-Y. Yeung, and Y. Xiong, "Super-resolution through neighbor embedding," in Proc. IEEE Conf. Comput. Vis. Pattern Class., 2004, vol. 1, pp. 275–282.
18. Kim, K.I., Kwon, Y.:Single-image super-resolution using sparse regression and natural image prior. IEEE Transactions on Pattern Analysis and Machine Intelligence 32(6), 1127–1133 (2010).
19. Bevilacqua, M., Roumy, A., Guillemot, C., Morel, M.L.A.: Low complexity single-image super-resolution based on nonnegative neighbor embedding. In: British Machine Vision Conference(2012).
20. Yang, J., Wright, J., Huang, T.S., Ma, Y.: Image super-resolution via sparse representation. IEEE Transactions on Image Processing (11), 2861–2873 (2010)
21. R. Timofte, V. De, and L. V. Gool. Anchored neighborhood regression for fast example-based super-resolution. In ICCV,2013.
22. R. Timofte, V. De Smet, and L. Van Gool. A+: Adjusted anchored neighborhood regression for fast super-resolution. In ACCV, 2014
23. J. Yang, J. Wright, T. S. Huang, and Y. Ma. Image super resolution via Sparse representation. ,TIP, 2010.
24. C. Dong, C. C. Loy, K. He, and X. Tang. Image superresolution using deep convolutional networks. TPAMI, 2015.
25. J. Kim, J. Kwon Lee, and K. M. Lee. Accurate image super resolution using very deep convolutional networks. In CVPR,2016.
26. Timofte, R., De Smet, V., Van Gool, L.: Anchored neighborhood regression for fast example-based super-resolution. In: IEEE International Conference on Computer Vision. pp. 1920–1927 (2013).
27. Yang, J., Lin, Z., Cohen, S.: Fast image super-resolution based on in-place example regression. In: IEEE Conference on Computer Vision and Pattern Recognition. pp. 1059–1066 (2013).
28. Chang, H., Yeung, D.Y., Xiong, Y.: Super-resolution through neighbor embedding. In: IEEE Conference on Computer Vision and Pattern Recognition (2004).
29. Rudin L I, Osher S, Fatemi E. Nonlinear total variation based noise removal algorithms. Proceedings of the 11th International Conference of the Center for Nonlinear Studies on Experimental Mathematics: Computational Issues in Nonlinear Science: Computational Issues in Nonlinear Science. Elsevier North-Holland, Inc. 1992:259-268.
30. Getreuer P. Image Interpolation with contour stencils. Quality of Life Research, 2011, 1(4):389-399.
31. D. Martin, C. Fowlkes, D. Tal, and J. Malik. A database of human segmented natural images and its application to evaluating segmentation algorithms and measuring ecological statistics. In ICCV, 2001.
32. C. G. Marco Bevilacqua, Aline Roumy and M.-L. A. Morel. Low-complexity single-image super-resolution based on nonnegative neighbor embedding. In BMVC, 2012.
33. R. Zeyde, M. Elad, and M. Protter. On single image scale-up using sparse-representations. In Curves and Surfaces, pages 711–730. Springer, 2012.
34. Huang J B, Singh A, Ahuja N. Single image super-resolution from transformed self-exemplars[C]// Computer Vision and Pattern Recognition. IEEE, 2015:5197-5206.
35. Schulter S, Leistner C, Bischof H. Fast and accurate image upscaling with super-resolution forests[C]// Computer Vision and Pattern Recognition. IEEE, 2015:3791-3799.
36. Dong C, Chen C L, Tang X. Accelerating the Super-Resolution Convolutional Neural Network[C]// European Conference on Computer Vision. Springer International Publishing, 2016:391-407.
37. Y. Wang, X. Lin, L. Wu, W. Zhang, Q. Zhang. LBMCH: Learning Bridging Mapping for Cross-modal Hashing. ACM SIGIR 2015.
38. L. Wu, Y. Wang. Robust Hashing for Multi-View Data: Jointly Learning Low-Rank Kernelized Similarity Consensus and Hash Functions. Image and Vision Computing, 57:58-66,2017.